# Moveable objects and applications, based on them

The inner views of all our applications are predetermined by the designers; only some non-significant variations are allowed with the help of adaptive interface. In several programs you can find some moveable objects, but it is an extremely rare thing; used for very specific things. However, the design of applications on the basis of moveable and resizable objects opens an absolutely new way of programming; such applications are much more effective in users' work, because each user can adjust an application according with his purposes. Programs, using adaptive interface, only implement the designer's ideas of what would be the best reaction to any of the users' doings or commands. Applications on moveable elements do not have such predetermined system of rules; they are fully controlled by the users. This article describes and demonstrates (with the help of the additional program) the new way of applications' design.

## Introduction

Imagine the situation that somewhere around the year 1750 you try to explain to the colonists of New England that there is an easy way to travel from Boston to Albany in less than three hours. Let's assume that you would be not prosecuted as the Devil's representative or declared a weak-headed. What would be the only way to prove that you are right? Only to take from the bushes some kind of a car and try to show, how it can move. My "car" is on www.sourceforge.net in the project **MoveableGraphics** (names are case-sensitive there!). Download from there the **Test_MoveGraphLibrary.zip** (to use from it the **Test_MoveGraphLibrary.exe** together with the **MoveGraphLibrary.dll**) or the **TuneableGraphics.exe**. This is the only way for you to see that whatever is described in this article is working, and maybe even better, than you'll expect from this text.

This is not my first article about the moveable / resizable screen objects. There are several articles, which demonstrate step by step the improvement of my algorithm, its use for simple and complicated objects, and its use in absolutely different areas and for different purposes [1 – 4]. The latest of these articles [4] was written only several months ago, but immediately after it I revised the foundations of my algorithm, which led to a more effective and more interesting basic level. With this new basis, nearly all the demonstration samples, which were used for explanation of all the different aspects, had to be rewritten, some samples disappeared, because they were not needed any more, but the new samples were developed for explanation of the new possibilities. The newest version is available at [6].

This article consists of the two main parts: the first part is about the algorithm, which allows to turn the screen objects into moveable / resizable; the second part is about the new type of programs – *user-driven applications* – which can be designed on the basis of such elements.

During the last 20 years after Windows conquered the world, people got used to some moveable objects on the screen. I am talking not about the moving of the objects according with some algorithm, which simply changes the coordinates of an object and redraw it at a new location. I am writing only about such movement of an object, which is not predetermined by a developer, but is decided only by a user. The type of movement / resizing, when a user can press an object at any point and move it to the new place or rotate, or a user can press at some border point and resize (reconfigure) an object.

Windows and similar operation systems turned a movement of rectangular windows into a common thing: any user can press the window's title bar with a mouse and move it to a new location, or press a border and resize this window. At the upper level we do it all the time, and we can't imagine a Windows system without such a thing. We can't imagine a system, in which not we, but the system itself, would decide about the position and size of each window. However, this is exactly the way, in which all the applications are working. We, users, can't decide about the inner view of the applications; this view is absolutely defined by the designer's ideas and is fixed at the development stage. From time to time we are allowed to make some changes, but only via some kind of selection, again predetermined at the development stage. The reaction on each of our choices is predetermined; this is the basic idea of any implementation of adaptive interface. The dynamic layout works in the same way, only the decisions are based on the font characteristics and the sizes of the window.

Maybe you can remember one of those rare programs that demonstrate some moveable elements. There is a couple of methods to turn some elements into moveable, for example, through painting on a panel, which is itself the subject of the dynamic layout. Such solution is very limited, but it is used from time to time. It is an extremely rare thing, when an element is really moveable; each time it requires a unique algorithm, taking into consideration the specific shape of an object and the way of its possible movement. And implementation of such an algorithm requires a very high programming level of its designer, so samples of such applications are really unique.

For many years I work on design of very complicated programs in different areas of science and engineering. Such programs are among the most sophisticated programs of all, and you would expect that they must change according with the progress of the computers and programming. In reality it's not so, and the best our-day scientific programs are nearly the same as they were 15 years ago. The process of design changed a lot (C# is much-much better for development than even

C++), but the results are the same; some strips and buttons are changed according with the new fashion, but applications are mainly the same. What is the cause of this stagnation? The main idea of the design: the applications are absolutely designer-driven.

Every designer has his level; it can be good or very good, but users of an application has no chances to go out of this level; they have always to work inside the designer's understanding of the problems. Users of the scientific / engineering applications are often much better specialists in the area, than the designers of the program, and yet they have to work under the level of the lesser specialist. Paradox, nonsense, but it is the law for all the designer-driven applications.

The abnormality of the situation became obvious many years ago, but to solve the problem, researchers and programmers always looked for new and new ideas in the area of adaptive interface. It's exactly like trying to fool the law of energy conservation and construct a *perpetuum mobile* by using the new materials and better lubrication. The basis of any adaptive interface is the fixed system of the reactions on any of users' doings (commands). This system of the predetermined reactions is coded by the developer, so the users of the applications with adaptive interface principally can't go beyond the designer's level of understanding the problems.

Several years ago it became obvious to me that until all the screen elements in the programs can be moved and resized only by developer of this program, users will have to stay inside the designer's understanding of what is the best for them. The only chance to bring the scientific / engineering applications to another level was in making all those elements moveable / resizable by users and in giving users the full control of the applications. When I started to work on this problem, I was looking for a solution that would allow:

- to turn into moveable / resizable the objects of any shape;
- to do it easily.

## An algorithm

My idea of making graphical objects moveable and resizable is based on covering an object with a combination of sensitive areas, some of which are used to start the forward movement (and/or rotation) of the whole object, others – for moving of some part(s), which cause either resizing or reconfiguring. The screen objects can be of an arbitrary shape, so I was looking for a minimal set of elements, which would give me a solution in any possible case.

A set of sensitive areas that covers an object, is called a *cover* (the `Cover` class). Each elementary sensitive area is called a *node* (the `CoverNode` class). The size of the nodes can vary – it is one of their parameters, but the possible shapes are restricted (in the currently working version of my algorithm) to such variants:

- <u>Circular</u> node, which is defined by its central point and radius.
- <u>Polygon</u> node, which is described by an array of points. <u>Polygon must be convex</u>!
- <u>Strip</u> node, which is defined by two points and radius. These points are the centers of semicircles at the ends of the strip; the strip's width is equal to the diameter of these semicircles.

The previous version of this algorithm was based on the different set of the basic elements. I switched to this newer version, when I found some samples, which had not the best solution for moving / resizing with the older variant. The consequences of turning all the screen elements into moveable / resizable are definitely much more important than the set of basic elements.

The moving / resizing of any object are defined by its `Cover`, which consists of an array of `CoverNode` objects. The number of nodes and their types are not important; any cover can be designed in many different ways, the nodes can overlap or stay apart. The order of the overlapping nodes can be very important as they may have different effect on moving / resizing, and they are checked for moving according with their order in the array. Usually the number of nodes is small even for really complicated objects, but there are cases, when a huge number of nodes are needed even for an object of a simple form.

Nodes have several parameters, which are very important for organizing the moving / resizing process. Any node gets the type of its possible individual movement, which is defined by one of these values { `None, NS, WE, Any` }. One of the node's parameters is the shape of a mouse cursor above this node. Each node of a cover has its personal number; when any node is caught or even sensed by a mouse, this number helps to identify the node. And this identification immediately tells about the possible movement of an object and determines the mouse cursor.

There are objects of two different types inside our applications – graphical objects and controls. Because of their principal differences, the covers for them are organized in different ways.

For turning graphical object into moveable / resizable, it must be derived from the `GraphicalObject` class, and three crucial methods must be overridden.

```
public abstract class GraphicalObject
{
    public abstract void DefineCover ();
    public abstract void Move (int dx, int dy);
    public abstract bool MoveNode (int i, int dx, int dy, Point ptMouse,
                                   MouseButtons catcher);
```
…

The cover of an object is described in the `DefineCover()` method. `Cover` class has a lot of different constructors, which can be used for design of covers for the real objects. In [6] you can find the whole **Test_MoveGraphLibrary** project with all the files in C# demonstrating a lot of different samples of cover design; in this article I'll mention only several of them.

`Move (dx, dy)` is the method for linear moving the whole object for a number of pixels, passed as the parameters.

The drawing of any graphical objects with any level of complexity is usually based on one or few very simple elements (`Point` and `Rectangle`) and some additional parameters (sizes). While moving the whole object, the sizes are not changed, so only the positions of these basic elements have to be changed.

`MoveNode (i, dx, dy, ptMouse, catcher)` is the method for individual moving of the nodes. The method returns a Boolean value, indicating whether the required movement is allowed; in the case of the forward movement, the `true` value must be returned if any of the proposed movements along the X or Y axes is allowed. If the movement of one node results in synchronous relocation of a lot of other nodes, it is easier to put the call to `DefineCover()` inside this method, and then it doesn't matter, what value is returned from the `MoveNode()`. This may happen for the forward movement, when movement of one node affects the relocation of the other nodes, and it usually happens with rotation, when all the nodes must be relocated.

Among the parameters for the `MoveNode()` method, there is a pair (`dx, dy`), which describes the forward movement, and there is a `ptMouse` parameter, which describes the position of the mouse cursor. It may seem that one of them is redundant and can be excluded, but I found out that the first one is excellent for the forward movement, and the exact mouse position is the best for organizing the rotation. A lot of the screen objects must be involved in both types of movements; the proposed algorithm works in any case. Covers are not designed differently for forward movement or rotation; there are no special parameters in nodes or covers, which would be especially for one or another type of movement, so any type of movement can be started at any place.

The whole moving / resizing process is organized for all types of objects only with a mouse. In all my applications, I use the left button to start the forward movement and the right button to start the rotation. This is not the law, but only the rule, which I try not to break. If you want to organize it in another way, you need only to make some minor changes in a code. Such samples are especially included into the **Test_MoveGraphLibrary** application and the needed changes are explained in [5].

The special features of controls prevent them from being turned into moveable / resizable in exactly the same way as the ordinary graphical objects. Controls have a specially designed system of reactions on all the mouse events, associated with their inner area; users are familiar with these reactions and expect them. Thus the only area, which can be used to start moving / resizing of controls, is the frame around their borders. If only a forward movement is needed, then such control can be moved by any point of a frame, which is close enough to its border; if the resizing is also expected, then the frame must be divided into areas, which are responsible for moving and resizing. All the standard controls have the rectangular shape, so users have no problems in finding those areas. To organize the moving / resizing of a control in the same way, as all other graphical objects, control is wrapped by a graphical object in the form of a frame. The area of such frame is divided into nodes, responsible for moving or resizing.

Giving the screen objects an ability to be moved and resized is only half of the design; there must be somebody to organize and supervise this whole process; in my algorithm it is called the `Mover` class. Further on I'll write mostly about the covers, their design and features, because these are the things that determine the possible movements of each object and what can be expected from the applications, based on such elements. Different design of covers gives programmers a chance to develop their applications in one way or another; the covers changes the way the objects are moved and resized. On the other hand, this supervisor class `Mover` is completely designed, and the whole process of moving / resizing is going according with the well known rules. Let's look at them in a quick way, because at some moments of the further explanation they are important.

- `Mover` supervises the whole moving / resizing process.  It doesn't matter how many moveable objects are there in the form, to what classes they belong, and how complicated are the movements of these objects; it's enough to have one `Mover` per form.  Though there is no such a strict rule, and there are situations, when it's better to use several `Movers`.

   ```
   Mover mover = new Mover ();
   ```

- `Mover` deals only with the covers and nothing else.  `Mover` doesn't know anything beyond the covers and its work never depends on any other object's parameters.

- `Mover` ensures the moving / resizing only for those objects that are registered with it (included into its List).

   ```
   mover .Add (…);
   mover .Insert (…);
   ```

- For the complicated objects, consisting of the parts, which can be moved both synchronously and independently, and for the objects, for which the set of such parts can be changed, while the application is running, it's much better to develop an `IntoMover()` method, which is used instead of manual registering of all the moveable parts and which guarantees the correct registering of an object and all its parts regardless of a set of constituents.

- Moving and resizing are done with a mouse, and the whole process is organized via the three standard mouse events: `MouseDown`, `MouseUp` and `MouseMove`.

- `MouseDown` starts moving / resizing by grabbing one of the nodes.  The only mandatory line of code in this method is

   ```
   mover .Catch (…);
   ```

- `MouseUp` ends moving / resizing by releasing any object that could be involved in the process.  The only mandatory line of code in this method is

   ```
   mover .Release ();
   ```

- `MouseMove` moves the whole object or a single node.  There is one mandatory line of code in this method, but in order to see the movement, the `Paint` method must be called

   ```
   if (mover .Move (mea .Location))
   {
       Invalidate ();
   }
   ```

Let's look at the covers of several objects, the shape of which is popular among the screen objects.  For better explanation, the figures in the text will show the sensitive nodes by additional lines and marks, but in the real applications these lines are never shown.  The whole idea of a good cover design is to organize the moving / resizing of the objects without any indication of those sensitive nodes.  Users have to know very few things:

- All graphical elements are moved and/or rotated by any inner point.
- All graphical objects are resized by borders.
- All controls are moved and resized, if allowed, by their frames.  The best places to resize the controls are their corners.

All the figures in this article are taken from the **Test_MoveGraphLibrary** application; for each sample only the name of the form will be mentioned.  All the objects in the mentioned application are moveable, so whatever is shown on the pictures is not exactly the view of those forms, when you open them, but what can be easily produced, when you move and resize the objects inside those forms.

### Covers for popular shapes

The most popular shape among the screen objects is a **rectangle**.

**Figure 1** shows four different types of possible rectangle resizing; these four types are defined by an enumeration.

   ```
   enum Resizing { None, NS, WE, Any };
   ```

The whole inner area of a rectangle is covered by a polygon node (in this case rectangular), which is used for moving of an object.  If a rectangle has to be resized, then its appropriate sides are covered by the narrow polygon nodes (again rectangular), which are used for resizing. If a rectangle needs to be resized in both directions, then not only all the sides are covered, but the circular nodes are added in the corners; the use of these nodes makes the resizing easier.

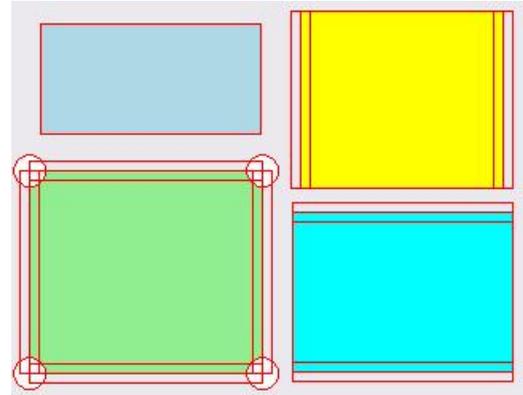

In the case of overlapping nodes, the moving / resizing depends on the order of nodes at the point. You can see at **figure 1** that there are areas, in which from two to four nodes can overlap. The priority of these nodes is set opposite to their sizes: the circular nodes in the corners have the highest priority, then go the nodes along the sides, and at the end of the queue is the biggest node for moving the whole rectangle.

I want to remind once more that in the real applications you'll never see these red rectangular frames or circles; there will be only normal objects, which can be moved and resized. This sample is from the **Form_RectangleGeneralCase.cs**, in which you can switch the visualizations of covers ON/OFF, and also change the sizes of all the nodes and try, how it works in different combinations. In that form, there is the `RectangleGeneral` class, whose cover is designed in the simplest way by using one of the standard `Cover` constructors.

**Fig.1** Covers for resizable rectangles

```
public override void DefineCover ()
{
    cover = new Cover (rc, resize, radius, halfstrip);
}
```

To organize those four moveable rectangles from **figure 1** with different types of resizing, four lines of code are used.

```
rg_None = new RectangleGeneral (rc None, Resizing .None, new RectRange (),
                                radius, halfstrip, Color .LightBlue);
rg_WE   = new RectangleGeneral (rcWE, Resizing .WE, new RectRange (),
                                radius, halfstrip, Color .Yellow);
rg_NS   = new RectangleGeneral (rcNS, Resizing .NS, new RectRange (),
                                radius, halfstrip, Color .Cyan);
rg_Any  = new RectangleGeneral (rcAny, Resizing .Any, new RectRange (…),
                                radius, halfstrip, Color .LightGreen);
```

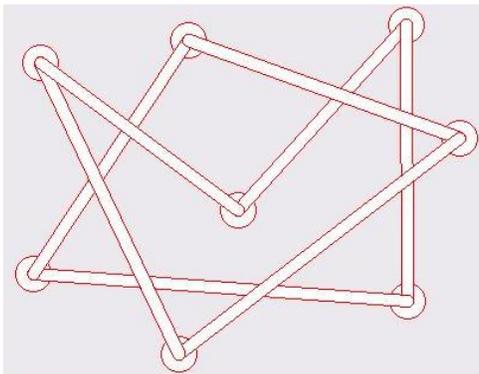

Another popular screen object is a **graph** – collection of points, connected with some lines. There can be different variants (rules) of organizing those connections; graphs of different types are used in absolutely different areas and tasks. **Figure 2** demonstrates a cover for a case, when a graph represents an infinitive loop.

The needed connections are represented by the strip nodes; the points are covered by the circular nodes. This design is used for a case, when each point can be moved individually, but by grabbing any connection, the whole object is moved. This sample is described in the **Form_InfinitiveLoop.cs**.

**Fig.2** Cover for an infinitive loop of points

It's very important to understand the difference between the image of a real object and the view of its cover. Cover is only a set of sensitive nodes to implement the needed moving / resizing operations. In the mentioned form, you can switch between the five different views, but the cover for all of them is the same. But it is also possible to achieve the same moving / resizing results by giving different covers to the objects of even the same class. There is no strict relation between a class and the type of cover for its objects. Objects of the same class may have different covers; objects of different classes may have identical covers.

Cover for an object from **figure 2** is also organized by using one of the standard `Cover` constructors.

```
public override void DefineCover ()
{
    cover = new Cover (pts, radius, halfstrip);
}
```

The use of different covers for the same objects, when they must be involved in different types of resizing, is demonstrated on the sample of the regular polygons. There are three different cases; all of them are in the **Form_RegularPolygons.cs**.

The first case of the moveable, but not resizable regular polygons can be seen at **figure 3**. As I mentioned before, one basic type of nodes is a convex polygon. In the case of rectangular objects, the nodes of the rectangular shape were used (**figure 1**), but this is only a private case of the polygon nodes. The only restriction is that the polygon must be convex, but there are no limitations on the number of apexes or the exact shape. Any regular polygon can be covered by a single polygon node; the only needed thing is the calculation of those apexes.

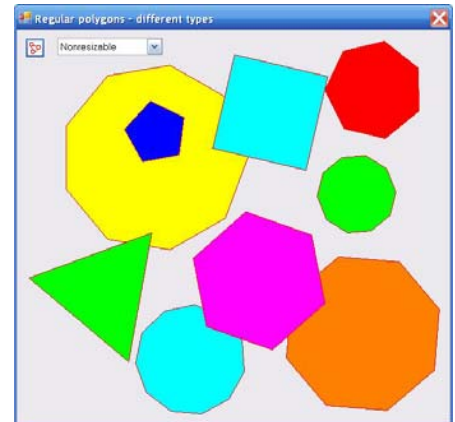

```
public override void DefineCover ()
{
    cover = new Cover (new CoverNode []
                { new CoverNode (0, Apexes) });
}
```

**Fig.3** Nonresizable regular polygons

The next case is of the regular polygons that can be zoomed by any apex (**figure 4**). The cover of such polygon also includes a big polygon node, covering the whole object's area, but it is preceded by a set of small circular nodes in the apexes.

```
public override void DefineCover ()
{
    CoverNode [] nodes = new CoverNode [nApexes + 1];
    PointF [] pts = Apexes;
    for (int i = 0; i < nApexes; i++)
    {
        nodes [i] = new CoverNode (i, pts [i], 5);
    }
    nodes [nApexes] = new CoverNode (nApexes, Apexes);
    cover = new Cover (nodes);
}
```

The third type of regular polygons can be zoomed by any border point, so instead of the circular nodes in the apexes, there will be the strip nodes, covering each segment of the border (**figure 5**).

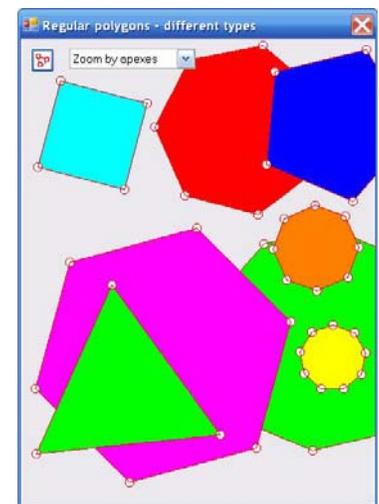

```
public override void DefineCover ()
{
    CoverNode [] nodes = new CoverNode [nApexes + 1];
    PointF [] pts = Apexes;
    for (int i = 0; i < nApexes; i++)
    {
        nodes [i] = new CoverNode (i, pts [i],
                pts [(i + 1) % nApexes], 5, Cursors .Hand);
    }
    nodes [nApexes] = new CoverNode (nApexes, Apexes);
    cover = new Cover (nodes);
}
```

**Fig.4**  Regular polygons, zoomed by any apex

The set of strips along the border looks similar to the cover of an infinitive loop from **figure 2**, only these polygons don't have circular nodes in the apexes. But the next sample will use even them.

A huge number of the screen objects can be represented by a polygon with an unlimited number of apexes. Some of these polygons are convex, others are not, but a polygon of an arbitrary shape can be divided into a set of triangles. Any triangle is a convex polygon; the triangles that degenerate into a line are excluded from consideration by other methods. **Figure 6** shows the transformations of polygons, represented by a set of triangles.

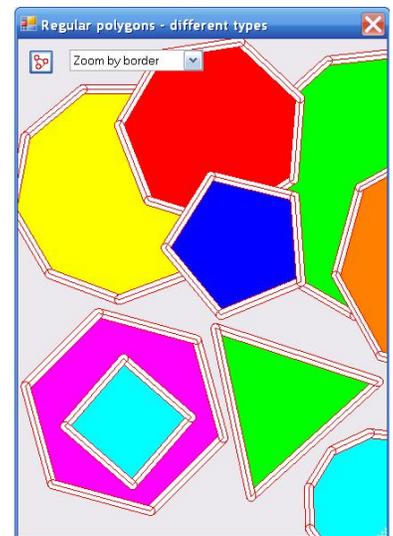

**Fig.5**  Regular polygons, zoomed by border

None of the chatoyant polygons at **figure 6** has a shape of the regular polygon, but each of them was born as a regular. The shapes, visible at this figure, are the results of different transformations. As one or another type of movement or transformation can be started at any point of an object, then the whole object must be covered by a set of nodes.

- Each apex is covered by a small circular node.
- The center point is also covered by a small circular node.
- Each border segment is covered by a strip node.
- The inner area is divided into a set of triangles; each of them covers an area between the center and two consecutive apexes.

The type of transformation depends on the part of an object, where it is grabbed with a mouse.

- Press any apex or center point and start the change of the shape.
- Press anywhere at the border and start zooming.
- Press anywhere inside and start forward movement or rotation (this depends on the pressed button).

It's only a coincidence that three different types of movements / transformations are started here with the different types of nodes. You can substitute the circles by the small squares (polygon nodes) and cover the border segments by rectangles (also polygon nodes); still all the movements can be organized in exactly the same way. As I mentioned before, the design of cover is not so important, if you receive the needed moving / resizing. Certainly, anyone would prefer to receive the needed result in the simplest way (with less and easier code).

As usual, all the movements are going regardless of whether the nodes are visualized (as on **figure 6**) or not. The switch from unicolor polygons to chatoyant makes the view of these figures more interesting and helps to find the central point without cover visualization.

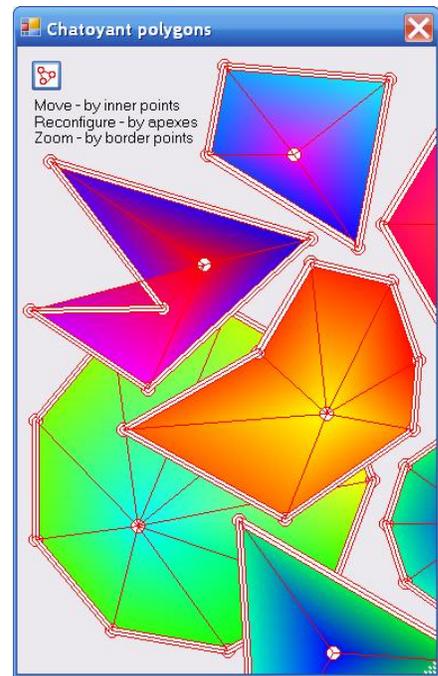

**Fig.6** Originally designed in the shape of a regular polygon, these objects can change their form

The special *N-node covers* are used for resizing of the objects with the curved borders. The N-node covers can be used in other cases, but that was the main purpose of their design: to cover the curved border with a narrow sensitive strip without any gaps, so that the objects with such borders can be resized by any border point.

**Figure 7** shows a ring with its N-node cover; such ring can be moved by any inner point and resized by any point of the inner or outer border. The cover consists of three different sets of nodes in such an order:

1. Circular nodes on the outer border.
2. Circular nodes on the inner border.
3. Polygon nodes between two borders.

The radius of the small nodes was selected in such a way as to make the resizing easy; the distance between the centers of the consecutive circles is determined by this radius so that the consecutive circles overlap. There is a couple of interesting situations, which can occur with the use of the N-node covers; they are explained in [5] on the samples of circles and rings.

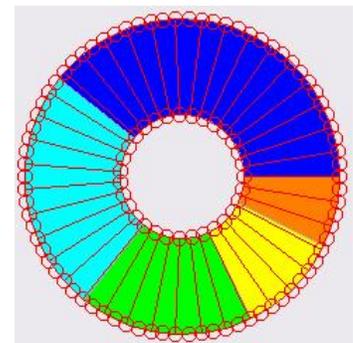

**Fig.7** A ring with an N-node cover

**Figures 1 – 7** demonstrate the covers for the graphical objects. All those objects can be moved by any inner point and resized (or reconfigured) by the areas of the nodes, especially included into the covers for such purpose (polygons from **figure 3** are declared nonresizable, and there are no such nodes). The situation with the controls is different. Though the standard controls have a rectangular shape, the covers from **figure 1** can't b used for them, because controls can't be moved by the inner points. I have already explained that controls receive a sensitive frame, parts of which are used for moving and other parts - for resizing. **Figure 8** shows variants of such cover; the sample is from the **Form_RectangleControlCase.cs**. There are different controls, used as samples, but each of them has some kind of textual information about the possible resizing of this control.

The control, to be moved, is wrapped in an object of the `ControlFrame` class. If no resizing is needed, then it is the simplest kind of a frame around an object, and the object can be moved by any point of this frame. In any other sort of resizing, there are circular nodes in all four corners, but depending on the exact type of needed resizing, these nodes have different types of their individual movements and types of shapes for cursor, when it is moved across them.

There are also additional nodes in the middle of those sides, in direction of which an object can be resized. In my applications, I prefer to use some correlation between the sizes of an object and the sizes of these nodes in the middle of the sides, so that it would be not a problem to find them even for really big controls. For the big controls, the nodes in the middle of the sides are enlarged, but even for the small controls there will be still some space between the corner nodes and the middle nodes. The covers, shown at **figure 8**, are never visualized in the real applications, in which the appearance of a node under the cursor (and thus the possibility of grabbing it to resize an object) is informed only via the change of the mouse cursor.

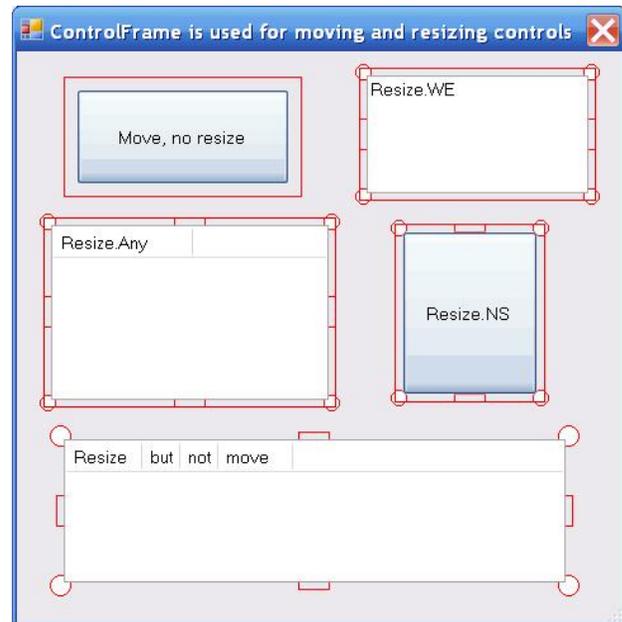

The radius of the corner nodes and the width of the frame are the parameters that can be set on declaring a control moveable / resizable. The enlarging of the circles over the width of the frame makes the finding of the corner nodes and the start of the resizing easier. The frame itself is always the last in the queue of nodes, so the possibility of resizing is checked before the possibility of moving.

The four upper controls at **figure 8** illustrate the four different resizing types: {None, NS, WE, Any}. The fifth sample (the lowest) is the only one without the solid red line around. This is a bit queer sample, but I want to show it, because it can be achieved by changing one parameter. This control is resizable, but not moveable in an ordinary way. Though it is moveable in the way all the cartepillars do it: move one side, then move the opposite side, and you'll be at a new place. Funny, but possible, though I never needed such a thing in a real application.

**Fig.8**  Different cases of moving / resizing controls

### From the simple covers to the most complicated applications

I have shown the covers for some of the most popular shapes of objects. These objects are simple and the covers for them are simple, but a very interesting fact that the covers for very complicated objects, used for design of the most sophisticated applications, are at the same level of simplicity. For example, let's look at one plotting area – an object of the `MSPlot` class, which is used in the scientific and engineering applications. An `MSPlot` object (**figure 9**) consists of:

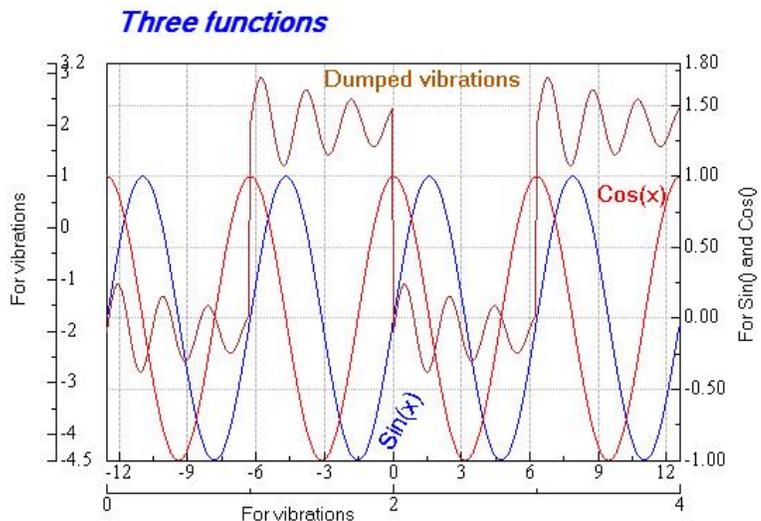

- One rectangular main plotting area.
- Any number of horizontal and vertical scales.
- An arbitrary number of comments, associated either with the main area or with the scales.

All these elements are moveable; some of them are also resizable. They are involved in different types of individual, synchronous, and related movements. Comments can be moved and rotated individually. Scales can be moved across the plotting area and positioned anywhere in relation to this area. When a scale is moved, all its comments move synchronously. Plotting area can be moved and resized. If it is moved, then all the scales and comments move synchronously; if it is resized, then the scales are resized also, and all the scales and comments try to keep their relative positions.

**Fig.9**  One plotting are with all its parts

As all these elements (comments, scales, and plotting area) can be moved, then all of them are derived from the `GraphicalObject` and have covers. The cover for any comment is a simple nonresizable rectangle, as was shown at **figure 1**. The cover for any plotting area is on the same **figure 1** as a case of the fully resizable rectangle. The cover for a scale would be also of a pure rectangular form, but… Scales can be moved and positioned anywhere in relation to their parent plotting areas. Very often they are positioned on the side of the plotting area, thus blocking this side from being used for resizing of the area. To leave a chance for resizing of the area even in the direction, where the side is closed by the scale, the cover of the scale gets "windows", through which the corners of the underneath area "look out", so a plotting area can be resized by all its corners regardless of the scales location.

**Figure 10** demonstrates the view of the **Form_GraphsAndComments.cs**; this is a typical view of any scientific application. The forms of such programs are populated with the plotting areas and controls. The plots are the complicated objects, which consist of several parts that can be involved both in individual and synchronous movements. Each part has its personal features, but the main rules are:

- whatever is moved, must be moved by any inner point;
- whatever is resized, must be resized by any border point;

These simple rules, applied to the objects on **figure 10**, results in such movements of the presented elements.

- **Plotting areas**  Are moved by any inner point and resized by any point on the border.
- **Scales**  Are moved by any inner point; resizing is done automatically with the resizing of the "parent" plotting area.
- **Comments**  Can be moved and rotated individually by any point. Each comment is associated with some plotting area or a scale; when such "parent" object is moved or resized, all its comments keep their relative position to the "parent".
- **Controls**  Are moved and resized (if needed) by the border.

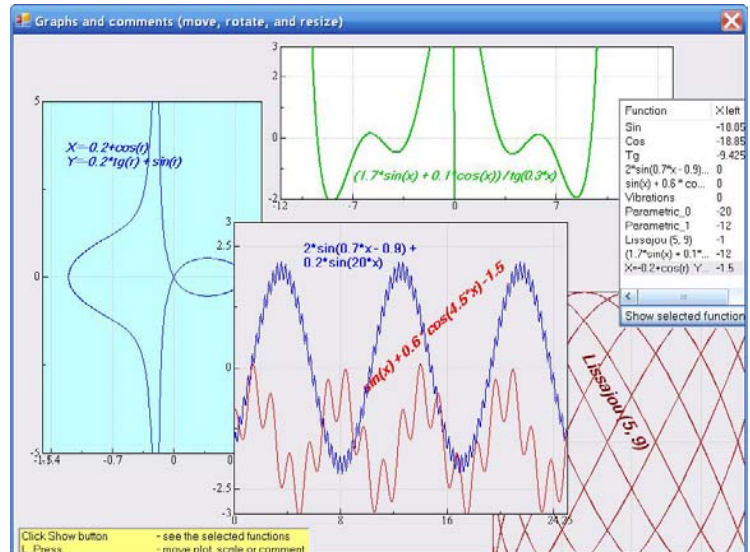

**Fig.10**  Typical view of a scientific application

The rules are simple and users get them in an instant. There can be any number of plotting areas with an arbitrary number of scales, comments, and graphs in each area. There can be a significant number of moveable objects on the screen; different context menus can be called for these objects; the objects can popup on top of others, or can be sent underneath. So many different things can be done here; all of them use an absolutely reliable system of identification, which guarantee the correct identification of any object under the cursor. Here are the main features of the scientific / engineering applications of such type.

- The forms are populated with the controls and different graphical objects (plotting areas, scales, and comments).
- Users decide about the placement of each and all parts. No limits on the number and size of the plotting areas.
- No limits on the number of comments; any number of them can be linked with each plotting area or scale.
- No non-moveable elements. Comments can be also rotated.
- Easy tuning of all the parameters. No restrictions on dealing with these tuning forms; all of them work independently, but if they work with the linked elements (like a plotting area and its scales), then they inform each other about the changes.
- Saving and restoring of all the visualization parameters. The number of tuneable parameters is huge; if a user spent some time on rearranging the view to whatever he prefers, then loosing of these settings is inadmissible. The easiness for a user of rearranging the whole view requires the system of storing the tuned areas somewhere in memory (in Registry) or in a file for using them later. Such system of saving / restoring is provided.
- A system of context menus covers a lot of requirements. Different menus are opened, depending on the place of its call; there are menus for plotting areas, for scales, for comments, and for an empty place. Partly these menus duplicate the actions that can be made in the tuning forms, but they are done much faster via the context menus. Other menu positions allow to do the unique things, which cannot be achieved in other ways.

There are no restrictions from designer; users can do whatever they need and want.

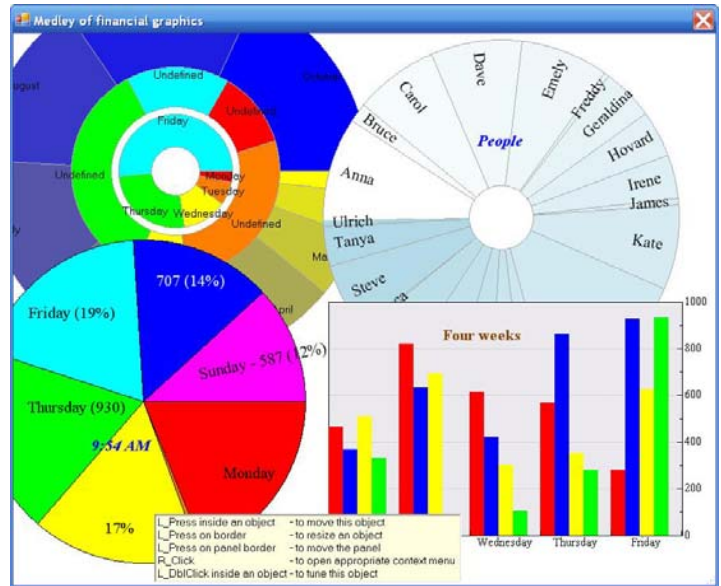

The next sample demonstrates the prototype of the user-driven application in the area of financial analysis; in many aspects the **Form_Medley.cs** (**figure 11**) is similar to the previous one.

- The number of objects is determined only by the user's wish; objects can be added or deleted at any moment via a context menu.

- Any object is moved by grabbing it at any inner point.

- Any object is resized by any of its border points. For a set of the coaxial rings this is applied also to the borders of each ring.

- Each class of financial graphics has its system of textual information; some of the texts are associated with the whole object, others – with the parts of objects. Each piece of textual information can be moved and rotated individually; it is also involved in synchronous movement with its "parent", when the last one either moved or rotated.

**Fig.11** An application for the analysis of the financial data

- All the visualization parameters, and there is a lot of them, can be easily tuned. The tuning can be done on an individual basis, or the parameters can be spread from one, used as a sample, on all the siblings, or the parameters can be spread to all the "children".

Though the applications, shown at **figures 10** and **11**, are from the absolutely different areas, they are designed under the same ideas and rules, common for all the user-driven applications. And the same rules produce outstanding results, when applied even to the forms, which consist mostly of the different types of controls.

Each type of scientific and financial graphics (only some of them are shown at **figures 9 – 11**), has a lot of visualization parameters, so each of them has one or more tuning forms. Such tuning forms in my applications and in all other numerous applications are designed on the same ideas. Usually the whole set of parameters is divided into several groups; parameters of each group are either linked with each other, or of the same type. Each parameter can be changed with the help of some control. Controls, associated with the parameters of one group, are positioned together and receive a short title, describing the idea of their grouping; very often such group also has a frame. The design of such tuning forms is also the nutrient medium for new ideas and articles in the area of adaptive interface. If you are an interface designer, you can constantly think out the new (definitely better!) placement of controls on the screen; if you work in the university and has to publish several papers a year, then those new surface layouts would be the source of inspiration long into your retirement. Turning the screen objects into moveable / resizable can greatly benefit the users, but can put an end to the articles about the best layout.

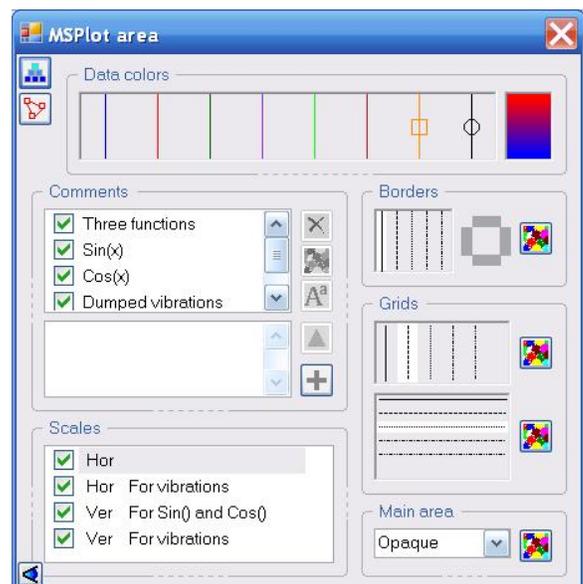

**Figure 12** demonstrates the tuning form for the main area of the `MSPlot` object, which was shown at **figure 9**. There are six groups in this form; each group has a frame with a title, explaining the main purpose of the group. Two groups have the standard frames; in other groups parts of the frames are changed into the dashed lines. This is not the most interesting feature of this form; the whole form behaves in such a way that you had never seen before.

A lot of parameters of the main plotting area can be changed in this form:

- Select the mode and the background color of the area.

- Select the type and color of the grids; select, which of them must be shown.

- Select the type and color of the borders; select which of four borders must be painted.

**Fig.12** Tuning form for an `MSPlot` main area (default view)

- Select the pens to draw the functions.
- Add / delete / modify / hide the comments of the main area.
- Hide or return back to view the scales.

All these tunings are needed from time to time, but obviously there are the parts, which are used much more often than others.  If a group is not used for a long time, it's better to take it out of view and reduce the valuable screen area, which is occupied by this form.  This is the classic of the adaptive interface, when you give users the list of items, users mark (select) the needed items, and others are taken out of view.  I designed such things years ago; you can find such interface in a lot of programs.  The instrument of selection doesn't matter at all, though the best way of organizing such a selection is discussed in many articles on adaptive interface.  The main thing here that the designer gives the predetermined list, the users select whatever they need; the resulting view after any kind of selection was coded by the developer long ago.  Adjusting this view to the forms sizes and the font, which user prefers, means that the developer is well familiar with the good practice of the dynamic layout.

The form at **figure 12** uses neither adaptive interface, nor dynamic layout.  Instead all the groups are turned into moveable; four of the groups are also resizable.  (From my point of view two other groups simply don't need any resizing.)  This switch to moveable groups allows to move them freely, position them in any way you want, and not needed groups can be simply thrown out of view.  The groups can be also resized; the dashed lines mark those sides of the frames that can change their length.  The available changes are more powerful than any type of the adaptive interface; all the changes can be done in the simplest way – with a mouse.  There is no sense in discussion, which view of the form is better or the best; any user can rearrange the form in any possible way and decide at any moment, how the form must look.

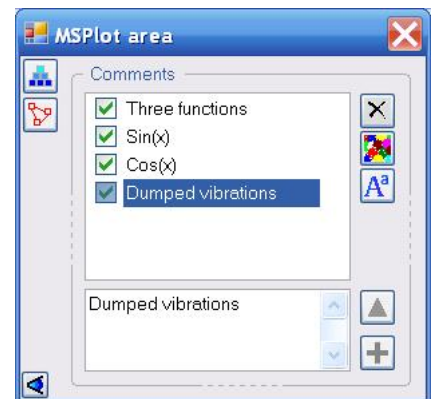

Suppose that you don't want to change any more the visualization parameters at **figure 9**, but you want to put different comments to the plots there, so you are interested only in the *Comments* group of this tuning form.  In a second you can rearrange it to **figure 13**.  At any moment there can be between one and six groups in view; the positions and sizes of these groups are not predetermined, but decided only by a user.

**Fig.13**   An `MSPlot` tuning form (customized view)

Though it is difficult to see anything common between the forms at **figures 10, 11** and **12**, they are designed under the same main rule: everything is moveable.  In the demonstration application, there is the **Form_PanelsAndGroups.cs**, which was especially designed to show different possibilities of moveable / resizable groups.  The group at **figure 13** and all six groups at **figure 12** belong to the `Group` class.  Objects of this class can be moved by any inner point and resized by any border point.  Thus the same basic rules that were used for graphical objects in the scientific and financial applications are applied to such widely used elements of design as groups, which include only different kinds of controls.

## Conclusion

The easy to use algorithm of turning any screen object into moveable / resizable changes not only the design of applications, but the use of them.  Applications, based on such elements, are turned from designer-driven into *user-driven*.  Each user gets a chance to work with an application that works at its best for any user at any moment.  It's not an adaptation of an application; this is a personally designed application.

**References**
1. S. Andreyev.  Design of moveable and resizable graphics. Cornell University Library, Computing Research Repository (CoRR), September 2007.
2. S. Andreyev.  Design and use of moveable and resizable graphics. Part 1.  In *Component Developer Magazine*, March/April 2008, pp. 58-69.
3. S. Andreyev.  Design and use of moveable and resizable graphics. Part 2.  In *Component Developer Magazine*, May/June 2008, pp. 56-68.
4. S. Andreyev.  Moving and resizing of the screen objects. Cornell University Library, Computing Research Repository (CoRR), September 2008.
5. S. Andreyev.  Moveable and resizable objects.  Currently revised version in the file **Moveable_Resizable_Objects.doc** at www.sourceforge.net in the project **MoveableGraphics.**
6. www.sourceforge.net   Project **MoveableGraphics** (name is case sensitive!).

Dr. Sergey Andreyev ( andreyev_sergey@yahoo.com )
April 2009